\documentstyle[aps,twocolumn,graphicx,colordvi]{revtex}
\include{pdflatex}
\begin{document}
\draft                       


\title{Physical properties of a GeS$_2$ glass using approximate \\ 
{\em ab initio}  molecular dynamics}
\author{S\'ebastien Blaineau and Philippe Jund}
\address{Laboratoire de Physicochimie de la Mati\`ere Condens\'ee , 
Universit\'e Montpellier 2, \\Place E. Bataillon, Case 03, 
34095 Montpellier, France}
\author{David A. Drabold}
\address{Dept. of Physics-Astronomy, Ohio University, 
Athens OH 45701, USA}

\maketitle

\begin{abstract}
With the use of {\em ab initio} based molecular dynamics simulations we
study the structural, dynamical and electronic properties of glassy 
g-GeS$_2$ at 
room temperature. From the radial distribution function we
find nearest neighbor distances almost identical to the experimental values
and the static structure factor is close to its experimental counterpart. 
From the Ge-S-Ge bond angle distribution we obtain the correct distribution
of corner and edge-sharing GeS$_4$ tetrahedra. Concerning the dynamical
characteristics we find in the mean square displacement of the atoms
discontinuous variations corresponding either to the removal of coordination
defects around a single particle or to structural rearrangements involving
a larger number of atoms. Finally we calculate the vibrational density
of states, which exhibits two well separated bands as well as some features
characteristic of the amorphous state, and the electronic density of states
showing an optical gap of 3.27 eV.
\end{abstract}
 

\section{Introduction}

Amongst the chalcogenide glasses, glassy germanium disulfide (g-GeS$_2$) 
has been heavily studied for many years \cite{weinstein} 
and is still the subject of recent experimental investigations 
\cite{salmon,boolchand} because of its interesting physical properties. 
Chalcogenide  materials can be used  as sensitive media for optical 
recording, as light guides, as high-resolution inorganic photoresistors 
or anti reflection coatings\cite{malek}. Moreover bulk glasses with 
for example Ag$^+$ cations are good  solid electrolytes with a high 
ionic conductivity at room temperature \cite{robinel}
and thin GeS$_2$ films are promising materials for submicron lithography 
when doped with silver \cite{hugget}. The high quantum efficiency 
of these glasses appears as a consequence of the relative high masses of the 
elements involved \cite{riseberg}. All these potential applications of 
glassy GeS$_2$  have led many authors to study the physical properties 
of these chalcogenide glasses and many experiments have been done on 
this topic \cite{exp,density}. 
However in order to understand the physical mechanisms occurring at the atomic
scale and leading to the results observed in experiments, numerical simulations
can be an alternative tool and more specifically Molecular Dynamics (MD) simulations. 
Although cluster modeling simulations were performed on g-GeS$_2$ \cite{cluster1}, it 
appears that GeS$_2$ compounds have not been the topic of extensive MD investigations yet, 
contrary to GeSe$_2$ \cite{refsall} or SiSe$_2$ \cite{vashishta}. In order to perform such 
investigations one has to decide what kind of description (classical or 
{\em ab initio}) is adequate for GeS$_2$. Taking the mostly (but not purely) covalent
bonding into account in g-GeS$_2$  a 
first-principles approach seems appropriate. In this paper we present 
therefore a theoretical study of the structural, dynamical and electronic 
properties of 
g-GeS$_2$ using an approximate {\em ab initio} description based on 
the Sankey-Niklewski scheme \cite{sankey} and materialized in the 
so-called ``FIREBALL96'' MD code \cite{fireball}. 
This technique has been successfully used in the study of several 
different chalcogenide glasses \cite{drabold,drabold2} and in order to 
check its validity in the case of GeS$_2$ 
samples we have compared our results with experimental results 
when those were available. 
Concerning the structure at 300K, the nearest-neighbor distances as well as 
the static structure factor compare well with the experimental data. Using 
the angle distributions and the radial pair distribution functions we 
find the correct proportion of edge and corner sharing GeS$_4$ tetrahedra 
which are the basic building blocks of the germanium disulfide glass.
Concerning the dynamics of the individual particles, we find in the mean 
square displacement (MSD) signatures of individual or collective 
atomic rearrangements corresponding to either the removal of ``defects'' 
or to the oscillation of large clusters which could be at the origin of the excess of modes 
seen at low frequency in the vibrational spectrum.\\
The paper is organized as follows: In section II we present the 
theoretical foundation of the FIREBALL96 code as well as the approximations 
used. In section III the results are presented for the structural, dynamical and electronic 
properties of the GeS$_2$ sample and section IV gives the major conclusions. 

\section{Model}

The theoretical framework of our work is the widely used Density Functional Theory (DFT) \cite{hohenberg-kohn} using {\em{three}} additional  
approximations. \\
First, we use the well known Local Density Approximation (LDA) 
\cite{ceperley} combined with  the pseudopotential approximation, which 
replaces the core electrons by an effective potential acting on the 
valence electrons (Hamman-Schluter-Chiang pseudopotentials 
are used \cite{schluter}). The electronic eigenstates are determined by a 
tight-binding-like linear combination of Pseudo Atomic Orbitals (PAO's) 
that satisfy the atomic self-consistent Hohenberg-Kohn-Sham 
equations \cite{kohn-sham}. 
A minimal basis set of one {\em{s}} and three {\em{p}} confined pseudo-orbitals per site is required. \\
The second approximation has been suggested by Harris \cite{harris}. 
It consists in using a sum of neutral-atom spherical charge densities as 
a 0$^{th}$-order approximation to the self-consistent density, keeping 
only the first-order changes from this density in the energy functional. 
This approximation avoids the necessity of iterating to self-consistency, 
so eigenvalues only need to be determined once instead of $\approx$10 times 
at each step. This approximation also avoids four-center Coulomb integrals 
in our calculations, which is a great simplification. 
The Harris functional has been used in many studies and has always 
given surprisingly good agreement with fully self-consistent 
calculations, except for highly ionic systems\cite{drabold,polatoglou}.\\
A third approximation is made to reduce the range of the 
tight-binding-like Hamiltonian matrix elements. To that purpose, 
the PAO's are slightly excited by imposing the boundary condition that 
they vanish outside a predetermined radius. This cut-off radius 
is chosen equal to $5a_0$, which represents a distance of 2.645 \AA. Atoms 
do not overlap each other beyond twice this distance, so the number of 
neighbors of each atom is considerably reduced.\\ 
All these approximations permit to gain a considerable amount
of CPU time compared to {\em ab initio} methods like the Car-Parrinello scheme \cite{car} and 
therefore one can perform longer simulation runs 
or study larger systems. Moreover this method has proved to be a very 
efficient tool for a wide variety of problems, and has been used with 
success in many different investigations \cite{drabold,drabold2,drabold3,ortega}.\\
Concerning the details of the present simulation, all of our calculations were 
performed in the microcanonical ensemble (N,V,E=constant), with a time step of 
2.5 fs, and using only the $\Gamma$ point to sample the Brillouin zone. 
The initial configuration of our system was a crystalline $\alpha$-GeS$_2$ 
sample containing 96 particles (32 Ge and 64 S) confined in a cubic cell 
of 13.82 \AA ~to which periodic boundary conditions have been applied. 
This represents a density of 2.75 g.cm$^{-1}$ , which is the usual 
experimental density \cite{density}. This crystalline configuration was then 
melted at 2000 K over approximately 2 ps and then equilibrated at 1000 K
for an additional 1.5 ps. We then quenched the system (by velocity
rescaling) through the glass transition (T$_g$=710 K) to a target temperature
of T=300K over 4ps (for more details on similar systems, see \cite{drabold2}).
Starting from this configuration, we performed a very long thermal
MD simulation at 300 K over 375 ps {\em i.e.} 150000 steps. During this
time, we saved the configurations every 20 steps and consequently
all the results presented below have been averaged over these
7500 configurations.

\section{Results}

\subsection{Structural properties}

The basic building blocks of glassy g-GeS$_2$ are GeS$_4$ tetrahedra, 
connected together within a random network. The structural unit disorder 
is reflected in the absence of long range order and in the wide 
distribution of bond lengths and bond angles. Structural information may 
be extracted from the radial pair correlation function $g(r)$.
For a given $\alpha - \beta$ pair it is defined by:
\begin{eqnarray}
g(r)_{\alpha - \beta}=\frac{V}{4 \pi r^{2} N_{\alpha} dr}~dn_{\beta}
\end{eqnarray}
Results are shown in Fig.~1 for the three different pairs. The smallest 
distance appears for the Ge-S pairs (Fig.~1(b)) at 2.22~\AA , and is
in perfect agreement with the distance determined experimentally (2.21~\AA \cite{exp}). 
The distance between two Ge atoms represents the intertetrahedral 
distance, and depends on the nature of the connection between the 
tetrahedra. The first peak at 2.91~\AA ~in Fig.~1(a) is due to 
edge-sharing tetrahedra, 
while the second one, at 3.41~\AA , is due to corner-sharing links as shown 
in Fig.~2. The experimental distances are respectively estimated at 
2.91~\AA ~and 3.42~\AA ~\cite{exp}, which is extremely close to our results. 
Finally the S-S pairs are responsible of the wide peak centered at 3.66~\AA 
~(Fig.~1(c)), which is also extremely close to the experimental first S-S 
distance of 3.64~\AA~\cite{exp}.\\
A complementary way to analyze the structure is to compute the static
structure factor $S(q)$ (obtained by Fourier transformation of $g(r)$)
which can be directly compared to its experimental counterpart.
In Fig.~3 we present the calculated $S(q)$ together with the one
obtained by neutron diffraction experiments \cite{bychkov}. The good 
agreement between the two curves shows the quality of the model concerning
the structural description of GeS$_2$ glasses.  The First Sharp Diffraction Peak (FSDP), 
which is a signature of the intermediate range order 
in amorphous states, appears at $\approx$1~\AA$^{-1}$ and is slightly 
underestimated compared to the experimental one. This is probably a 
consequence of the small size of our system: 1~\AA$^{-1}$ represents in
real space a distance of 6.3~\AA ~and a sphere with such a radius provides 
a volume which is close to the total volume of our cell. Therefore the lack 
of statistics for these large distances can explain the underestimation of 
the FSDP in our simulation.\\ 
In order to analyze completely the medium-range structure we have also
calculated the bond angle distributions and in particular the intratetrahedral $\widehat{SGeS}$ 
and intertetrahedral $\widehat{GeSGe}$ bond angles which are
represented in Fig.~4. The intratetrahedral angle $\widehat{SGeS}$ is 
centered at 110$^{\circ}$, which is close to the perfect tetrahedral 
angle of 109.47$^{\circ}$. Its large distribution is a signature of the 
structural disorder of our glassy sample. The intertetrahedral bond angle 
$\widehat{GeSGe}$ is the angle {\em between} tetrahedra and includes 
two major contributions. The first one, centered at 80$^{\circ}$, is 
caused by edge-sharing tetrahedra. The second, at $\approx$100$^{\circ}$, 
is due to corner-sharing tetrahedra. The integration of these two peaks 
permits to estimate the fraction of edge-sharing and corner-sharing connections 
which are respectively 18.6$\%$ and 81.4$\%$. These results have also 
been confirmed by a direct counting of each type of connection in our 
sample. Experimental Raman scattering measurements in amorphous g-GeS$_2$ have 
given 16.6$\%$ of edge-sharing links and 83.4$\%$ of corner-sharing links \cite{boolchand} 
which is relatively close to our results.\\
In view of all these data, we can safely say that the model describes
correctly the structure of amorphous GeS$_2$. It remains to be seen if
this is also true for the dynamical properties which is the topic of
the next section.

\subsection{Dynamical properties}

The dynamical properties of glassy g-GeS$_2$ have been studied through the 
Mean  Square Displacement (MSD) and the vibrational density of states (VDOS).
The MSD is defined as $\langle r^2(t)\rangle=\langle|\vec{r_i}(t)-\vec{r_i}(0)|
^2\rangle$ where $\vec{r_i}(t)$ is the position of particle $i$ at
 time $t$. We can deduce from the slope of the MSD the diffusion constant 
$D$, where  $D=\frac{1}{6}\lim_{t\to\infty}\frac{r^2(t)}{t}$ . 
In our calculation, $D$ was found equal to zero; this means that the thermal 
energy at ambient temperature is not high enough to reach the diffusive 
regime during the time scale of our simulation (375 ps). Nevertheless
during this time some specific structural rearrangements can occur 
which manifest themselves by a brutal increase of the total MSD or of
the MSD of individual atoms. In this later case the ``jumps'' in the MSD 
are due to the removal of a coordination ``defect'' in the glassy system.
An example of such a rearrangement is shown in Fig.~5: Fig.~5(a) represents
the individual MSD of Ge$^\star$, a particular germanium atom, with a
brutal increase from $\approx$~0.5~\AA$^2$ (before the jump) to 
$\approx$~2.5~\AA$^2$ (after the jump) around 100~ps. The jump can clearly be seen in
 Fig.~5(b) which shows the projection of the displacement of Ge$^\star$ 
on the x-z plane while the reason of the jump becomes apparent in 
Fig.~5(c) and 5(d) which illustrates the group of particles surrounding 
Ge$^\star$ just before 
and after the jump. Indeed we see that the initially 3-coordinated Ge$^\star$ 
atom gets linked with a terminal sulfur atom creating thus 2 edge-sharing 
tetrahedra which is a configuration energetically more favorable.\\
The second kind of rearrangement illustrated in Fig.~6 involves a larger 
number of particles and manifests itself by a ``pulse'' in the total MSD 
whose amplitude is more important for the sulfur atoms than for the
germanium atoms as shown in Fig.~6(a). In that case a group of 
particles ($\approx$~20) in a certain configuration at $t_1$ (Fig.~6(b)) 
switches to a new state at $t_2$ (Fig.~6(c)) which can be called metastable 
since its life time is relatively short ($\approx$~10~ps) before the 
system comes back again to its original structure. Note that in this case 
no link has been broken or created. We observed such ``oscillations'' twice
in our simulation with a time interval of 300~ps. Clearly our simulation
time is too short to see if these oscillations repeat themselves at a 
well determined (low) frequency and to make a connection with the
so-called ``soft''-modes \cite{parshin} well known in amorphous systems.\\
To complete the study of the dynamical properties we have computed $g(\nu)$,
 the Vibrational Density Of States (VDOS), via a Fourier transformation 
of the velocity autocorrelation function:
\begin{eqnarray}
g(\nu)=\frac{1}{Nk_b T}\Sigma_i m_i \int_{-\infty}^{\infty}exp(i\nu t)\langle\vec{v_i}(t).\vec{v_i}(0)\rangle dt
\end{eqnarray}

The Fourier transformation has been calculated using the Wiener-Kinchin theorem \cite{wiener} 
over the last 4096 steps of the simulation. 
The total spectrum as well as the partial contributions due to Ge and S are
shown in Fig.~7. Despite serious efforts we could not find the experimental
counterpart of the total spectrum since apparently no neutron diffraction 
studies have been performed on g-GeS$_2$. But comparing our results with
those obtained for the analogous GeSe$_2$ glasses \cite{drabold2} 
the spectrum exhibits the same features. Mainly two bands can 
be distinguished: a low-energy acoustic band involving mainly 
extended interblock vibrations and a high-energy optic band consisting of 
more localized intrablock vibrations.
The two main bands are clearly separated and have approximately the same
width (7~Thz).

In addition to the usual acoustic and optical bands, a  small band can be seen close to 8~Thz corresponding
to the so-called A$_1$ mode \cite{drabold}. The A$_1$ mode is well known to be a tetrahedral breathing mode
(in which a central Ge atom is stationary and its four S neighbors move radially relative to the
fixed Ge). This feature is strongly revealed in Raman measurements\cite{raman}, because the mode is 
especially Raman active. In Raman measurements, there is a clear
indication of a ``two peak" structure to the A$_1$ band.  In particular, one usually sees a high
frequency peak or shoulder which is interpreted as arising from edge-sharing tetrahedra (see Fig.~2), 
and the main band from tetrahedra in corner-sharing conformations\cite{sugai}. The $A_1$ and $A_{1c}$ modes 
have also been resolved  in inelastic neutron scattering studies of g-GeSe$_2$\cite{capdad}. In our work the 
tetrahedral breathing band does show a clearly resolved splitting. It is 
possible that a direct analysis 
of the eigenvectors of the dynamical matrix\cite{capdad} would provide more information linking the observed 
spectral feature to microscopic vibrational excitations. At the low 
frequency end of the spectrum, a shoulder is present between 
1 and 2~THz which is coherent with the existence of a ``Boson'' peak
found experimentally \cite{tanaka}. The ``Boson'' peak refers to an excess
in the VDOS with respect to the Debye distribution and is located 
generally around 1.5THz. 

\subsection{Electronic properties}

In Fig.~8 we present the electronic density of states (EDOS), obtained by 
binning the density functional electron energy eigenvalues from the 
starting, fully relaxed model. The $\Gamma$ point optical gap is 
3.27~eV which compares very well with the experimental value of 3.2~eV 
obtained by resonant Raman-scattering spectroscopy \cite{tanaka}. 
This good agreement is due to the opposite effects of the use of a minimal 
basis set which is well known to exaggerate the gap and of Kohn-Sham 
eigenvalues which tend to underestimate the gap. It should be noted that
this gap is greater than that (2.3~eV) of g-GeSe$_2$. Another point 
from Fig.~8 is the lack of any localized states in the optical gap 
(the Fermi level is at $E=0$ in our calculation). This lack of gap states is
realistic, since the density of gap states is very small in g-GeS$_2$. 

\section{Conclusion}

We find that the results obtained for our 96-atom GeS$_2$ model are in 
excellent agreement with all the corresponding experimental results 
that are available. This realism is surprising since the size of our 
system is relatively small, and accurate MD simulations usually require 
larger systems.\\
The structural properties of g-GeS$_2$, which have been extensively studied, 
are all extremely realistic in our simulation. The pair-correlation 
functions lead to interatomic distances that are within $10^{-2}$\AA ~compared
to the experimental values and the static structure factor is very 
similar to the one obtained from neutron diffraction studies. The small
underestimation of the FSDP encourages us to use larger models and we
are currently preparing samples containing 258 atoms.
The fraction of edge and corner-sharing tetrahedra, which can be 
deduced from the angular distribution, is also close to experiment. 
It should be mentioned that we don't find homopolar (Ge-Ge or S-S) 
bonds in the present investigation but their existence can not be excluded 
{\em a priori} in the 258-atom model. Probably a more in-depth study of the 
large system will permit to solve the apparent disagreement between 
two recent experimental studies on this point \cite{salmon,boolchand}.\\
Concerning the dynamical properties of our sample we find discontinuous
atomic displacements at ambient temperature, leading to jumps in the MSD.
These jumps can either be due to the removal of coordination defects
around a single atom or to oscillations of larger groups of atoms
($\approx$ 20) between a stable and metastable configuration which could
be at the origin of ``soft''-modes that are often seen in amorphous
systems. The vibrational density of states of glassy GeS$_2$ could 
not be compared directly to the experimental spectrum since to our knowledge 
it is not available in the literature yet. We find basically two 
bands separated by a ``gap'' in which exists a small structure due to the
tetrahedral breathing modes. At low frequency we find at around 1~Thz 
a shoulder corresponding to the famous Boson peak present in many 
amorphous systems. Concerning the electronic properties we find an
optical gap of 3.27 eV and no localized states in the gap which is 
in agreement with experimental data.\\
These first MD results obtained for g-Ge$_2$ with the use of the
Fireball96 code show that one can have confidence in this ``pseudo''
{\em ab initio} scheme giving an excellent description of the 
physical characteristics of germanium disulfide for a relatively
low computer load.\\

{\bf Acknowledgments:} PJ would like to thank R. Jullien, A. Pradel and 
M. Ribes for their help in initiating this project. We would like to 
thank also E. Bychkov for the data of the experimental structure factor 
of g-GeS$_2$ as well as Jun Li for advice in the
early stages of this work. Part of the calculations have been performed at 
the ``Centre Informatique National de l'Enseignement Sup\'erieur 
(CINES)'' in Montpellier. DAD thanks the National Science Foundation for 
support under grants DMR-0081006, 0074624 and 0205858.

\newpage

\begin{figure}
\centerline{\includegraphics[width=8.cm]{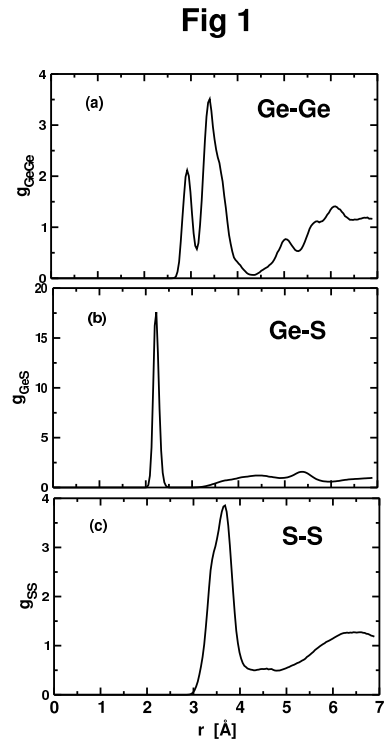}}
\caption{
Radial pair distribution functions: (a) Ge-Ge, \\(b) Ge-S, (c) S-S
}
\label{fig1}
\end{figure}
\newpage

\begin{figure}
\centerline{\includegraphics[width=10cm]{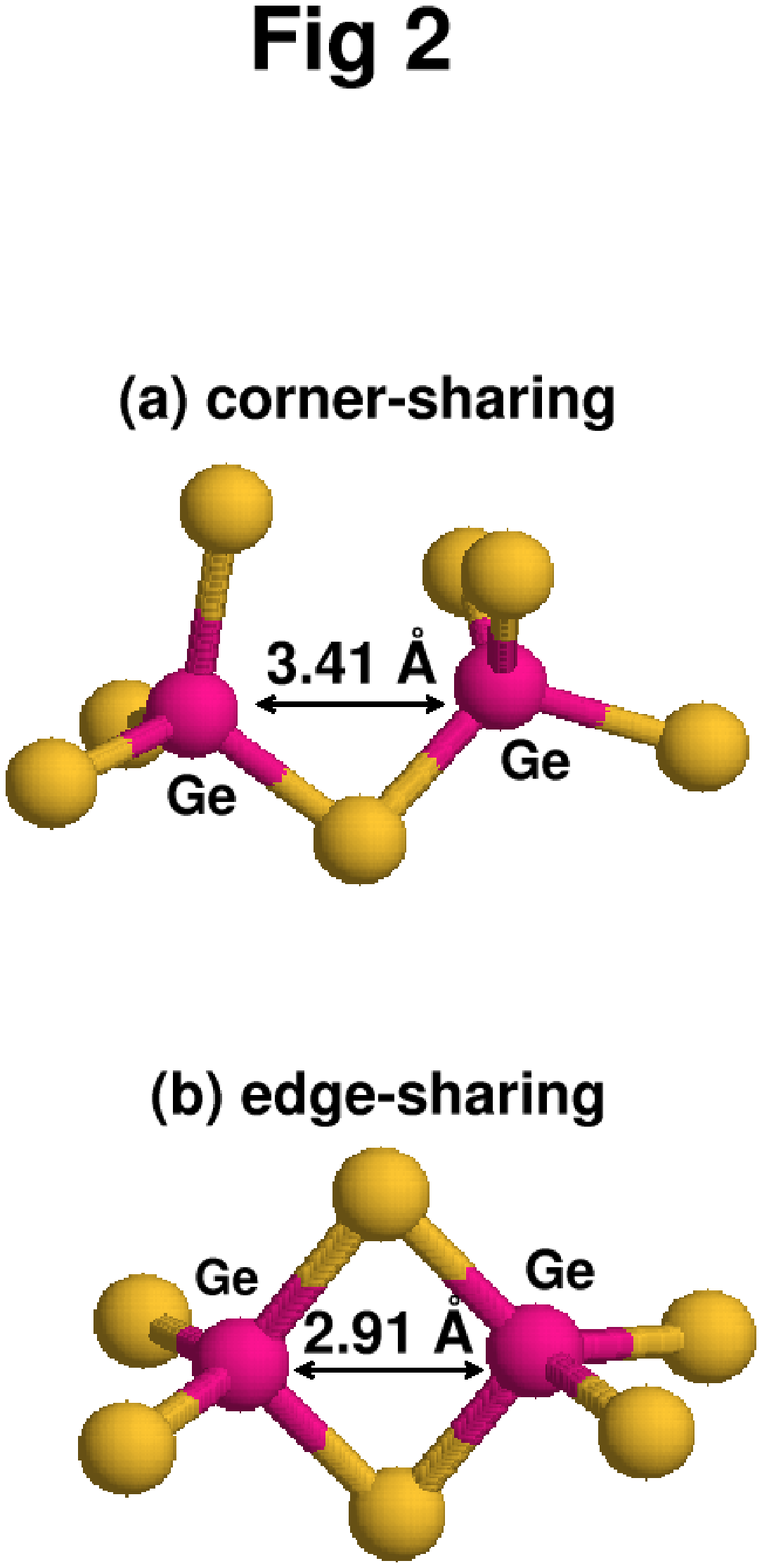}}
\vspace*{1cm}
\caption{
Ge-Ge distances in corner and edge-sharing tetrahedra
}
\label{fig2}
\end{figure}
\newpage

\begin{figure}
\centerline{\includegraphics[width=10cm]{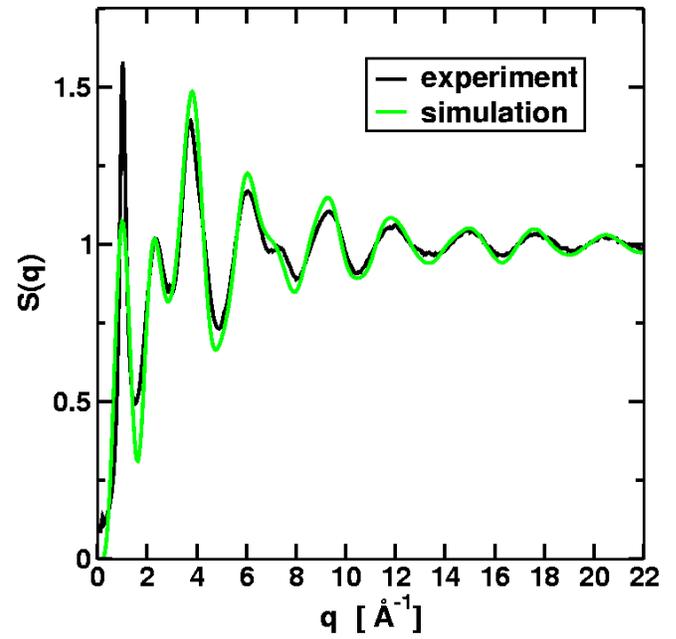}}
\vspace*{1cm}
\caption{
Experimental and simulated static structure factor
}
\label{fig3}
\end{figure}
\newpage

\begin{figure}
\centerline{\includegraphics[width=10cm]{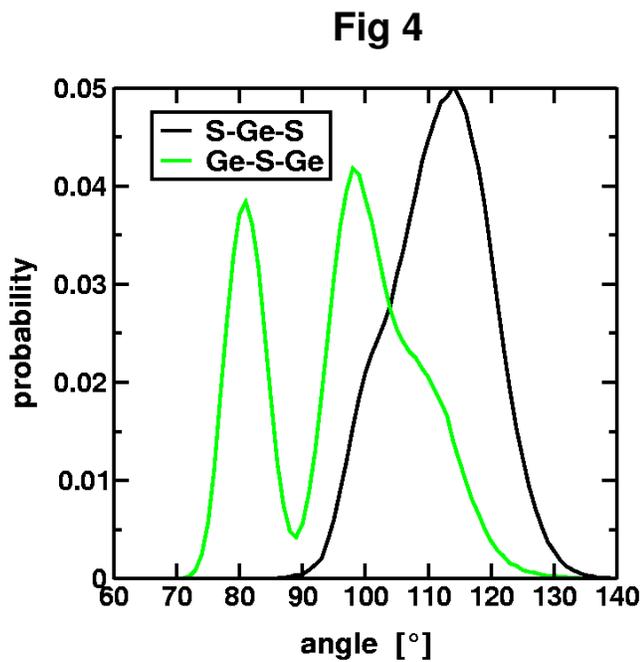}}
\vspace*{1cm}
\caption{
Bond angle distributions.
}
\label{fig4}
\end{figure}
\newpage

\begin{figure}
\centerline{\includegraphics[width=10cm]{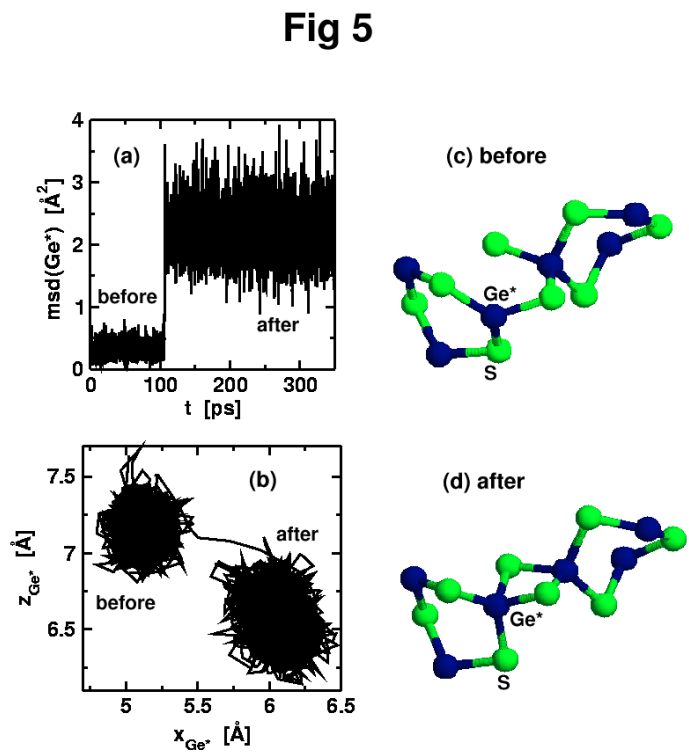}}
\vspace*{1cm}
\caption{
(a) Mean Square Displacement of Ge$^\star$ before and after the jump;
(b) Projection on the (x,z) plane of the trajectory of Ge$^\star$;
Atomic configuration around Ge$^\star$ before (c) and after (d) the jump.   
}
\label{fig5}
\end{figure}
\newpage

\begin{figure}
\centerline{\includegraphics[width=10cm]{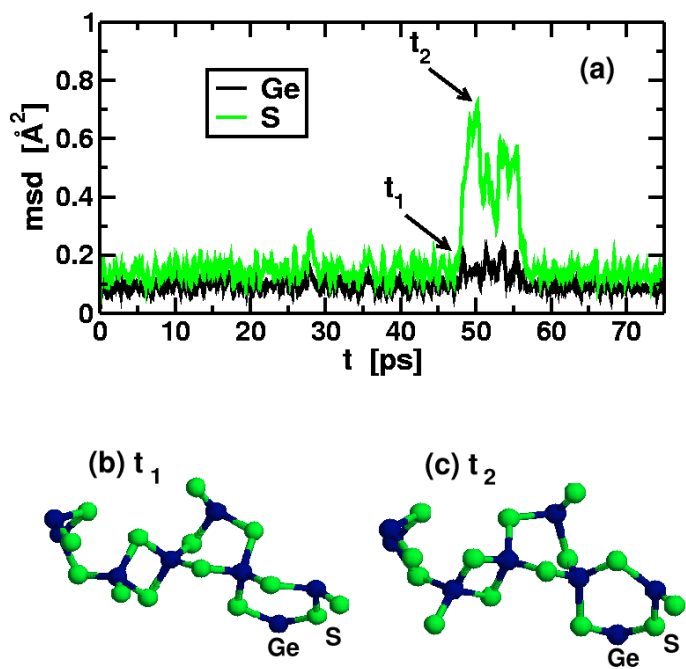}}
\vspace*{1cm}
\caption{
(a) Total MSD for the S and Ge atoms displaying a pulse at t$_1$.
Structural arrangement of the most mobile atoms at t$_1$ (b) and 
 t$_2$ (c)
}
\label{fig6}
\end{figure}
\newpage

\begin{figure}
\centerline{\includegraphics[width=8cm]{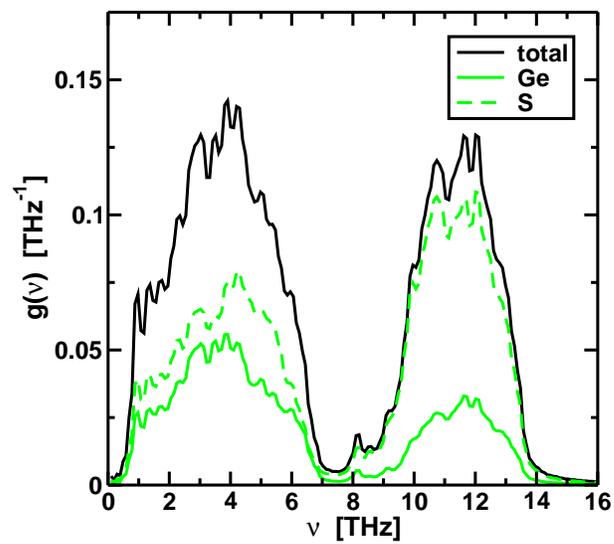}}
\vspace*{1cm}
\caption{
Total and partial vibrational density of states
}
\label{fig7}
\end{figure}

\begin{figure}
\centerline{\includegraphics[width=12cm]{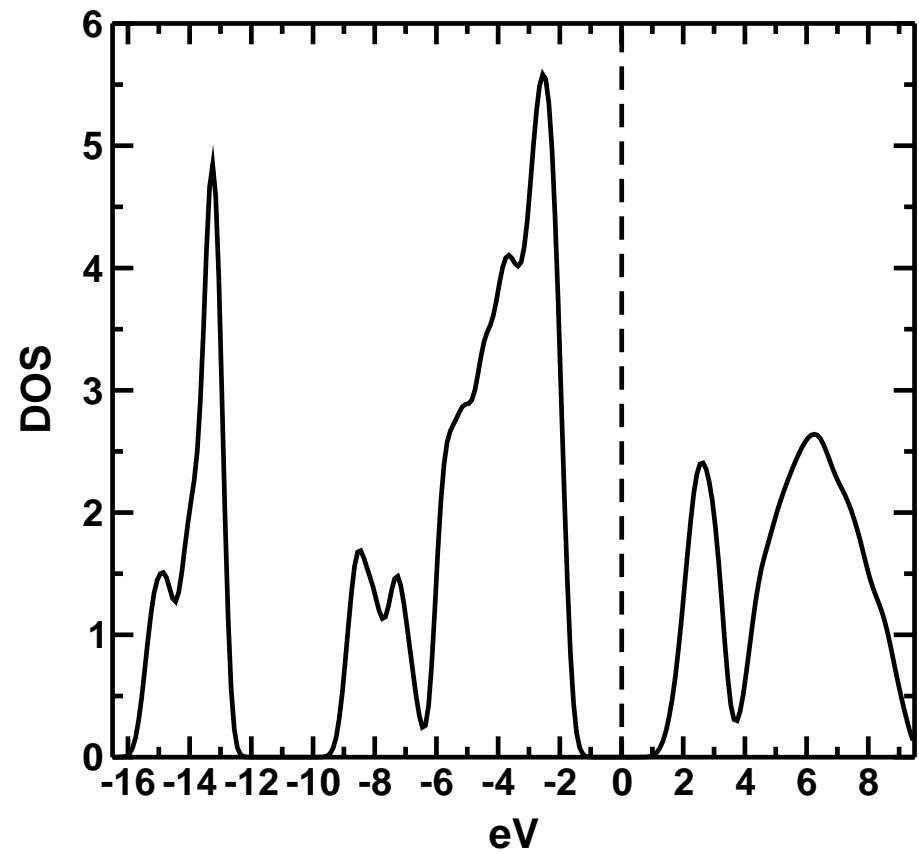}}
\vspace*{1cm}
\caption{
Electronic density of states. Dashed vertical line is the Fermi energy. 
The optical gap is found to be 3.27 eV.
}
\label{fig8}
\end{figure}


\end{document}